\documentclass{jfm}
\usepackage{epstopdf, epsfig}

\usepackage[caption=false,labelformat=simple]{subfig}
\usepackage{floatrow}
\usepackage{multirow}
\newcommand{\pardt}[1]{\frac{\partial #1}{\partial t}}

\newcommand{\pardr}[1]{\frac{\partial #1}{\partial r}}

\newcommand{\esth}[1]{\langle e_{#1} \rangle}

\newcommand\ddfrac[2]{\frac{\displaystyle #1}{\displaystyle #2}}
\newcommand\Ri{\mbox{\textit{Ri}}}  

\newcommand{\drawline}[2]{\raisebox{2.5pt}{\vbox{\hrule width #1 pt height #2 pt}}}
\newcommand{\spacce}[1]{\hspace{#1pt}}
\newcommand{\solid}{\drawline{24}{0.5}\spacce{2}}
\newcommand{\ccline}{\drawline{5.5}{0.5}\spacce{2}\drawline{1}{0.5}\spacce{2}}
\newcommand{\dline}{\drawline{5.5}{0.5}\spacce{2}}
\newcommand{\dashdot}{\nobreak\mbox{\ccline\ccline\dline}}

\usepackage[final]{changes} 

\shorttitle{Bifurcation of gyrotactic suspension}
\shortauthor{L. Fung and Y. Hwang}

\title{A sequence of transcritical bifurcations \\ in a suspension of gyrotactic microswimmers in vertical pipe}

\author{Lloyd Fung\aff{1}
  \corresp{\email{lloyd.fung@imperial.ac.uk}}
 \and Yongyun Hwang\aff{1}}

\affiliation{\aff{1}Department of Aeronautics, Imperial College London, London, SW7 2AZ, UK}

\begin{document}

\maketitle

\begin{abstract}
Kessler (\textit{Nature}, vol. 313, 1985, pp. 218-220) first showed that plume-like structures spontaneously appear from both stationary and flowing suspensions of gyrotactic microswimmers in a vertical pipe. Recently, it has been shown that there exist multiple\deleted{number of} steady, axisymmetric \added{and} axially uniform solutions to such a system (Bees, M. A. \& Croze, O. A., \textit{Proc. R. Soc. A.}, vol. 466, 2010, pp. 2057-2077). In the present study, we generalise this finding by reporting that a countably infinite number of such solutions emerge as Richardson number increases. Linear stability, weakly nonlinear and fully nonlinear analyses are performed, revealing that each of the solutions arises from the destabilisation of \added{a} uniform suspension. The \replaced{countability}{discrete number} of the solutions is due to the finite flow domain, while the transcritical nature of the bifurcation is because of the cylindrical geometry which breaks the horizontal symmetry of the system. It is further shown that there exists a maximum threshold of achievable downward flow rate for each solution if the flow is to remain steady, as varying the pressure gradient can no longer increase the flow rate from the solution. \added{All of the solutions found are unstable,} except \added{for} the one arising at the lowest Richardson number, \deleted{all of the solutions found are unstable,} implying that they would play a role in the transient dynamics in the route from a uniform suspension to the fully-developed gyrotactic pattern. 
\end{abstract}


\section{Introduction\label{sec:introduction}}

Gyrotaxis describes the biased swimming of bottom-heavy micro-organism, such as \emph{Chlamydomonas} and \emph{Dunaliela}, under the combined influence of ambient vorticity and gravity. It is a result of the balancing between the viscous torque originating from the vorticity and the gravitational restoring torque from the bottom heaviness of the cell. Gyrotaxis has been shown to be an important mechanism responsible for\deleted{ the} phenomena such as bioconvection \citep{Pedley1988,Bees2020} and gyrotactic trapping of algal species in the ocean \citep{Durham2009}.

The term `gyrotaxis' was first coined by \cite{Kessler1984, Kessler1985b, Kessler1985a, Kessler1986}. In \replaced{a}{the} series of experiments, he was able to single out the gyrotactic mechanism by observing an initially uniform suspension of bottom-heavy motile cells, which undergoes gyrotactic focusing along the axis of a long downflowing pipe. \cite{Kessler1986} proposed a simple deterministic model to describe the gyrotactic plume, \replaced{in which}{where} he assumed that the net flux of cells swimming towards the centre is proportional to the local shear rate and that the diffusivity is constant and isotropic.
Using the same model, \cite{Pedley1988} analysed the linear stability of a uniform and stationary suspension of gyrotactic swimmers in \added{an }infinite spatial domain. In the uniform suspension where the mechanism of gravitational overturning in thermal convection \citep{Childress1975} is absent, they found that gyrotaxis itself causes an instability (i.e. gyrotactic instability), which is likely to be responsible for the plume formation in deep suspensions \citep{Kessler1985a}. \cite{Pedley1988} further compared the wavelength of the plume structures predicted by the linear stability analysis with that observed in experiments, but failed to obtain \added{a }good agreement. They proposed three potential origins for the discrepancy: 1) finite-depth effect; 2) poor estimation of diffusivity; 3) nonlinear evolution of the plumes. The first issue was tackled by \cite{Hill1989}, and later by \cite{Bees1998}, where the linear stability of shallow suspensions was examined. The second has previously been discussed extensively \citep[etc.]{Bearon2012,Croze2013,Croze2017}, and recently by \cite{JFM2020} and \cite{Jiang2020}. In particular, these \replaced{works}{work} demonstrated that the generalised Taylor dispersion (GTD) model provides a more accurate description for the dispersion and instability of the suspension.

The objective of this work is to address the third issue, i.e. the role of nonlinearity in the plume formation. In particular, we shall consider a vertical pipe with \deleted{flow }both upward and downward \added{flow }and exhaustively seek the nonlinear solutions. For simplicity, here an infinitely long pipe is considered and the solution is assumed to be uniform in the axial and azimuthal directions.
Several previous \replaced{studies}{work} have \replaced{considered}{studied} the steady solutions for the nonlinear axially-uniform gyrotactic plume in a downflowing pipe.
\cite{Kessler1986} first derived an analytical solution, but it was limited to the case where the pressure gradient is zero.
\cite{Bees2010a} obtained the solution asymptotically and found that there can be more than one solution for a given set of parameters.
More recently, \cite{Bearon2012} and \cite{Croze2013,Croze2017} have computed the solution numerically using the GTD model to demonstrate its outperformance against the earlier model by \cite{Pedley1990}.

Despite these \replaced{advances}{progresses} made over the years, little attention has been paid to the role of nonlinearity in the existence of the solutions itself. Indeed, it has only been very recently \added{discovered} that the solution for \added{a} uniform and stationary suspension undergoes a transcritical bifurcation, through which a non-trivial solution emerges in the form of a single vertically uniform plume \citep{JFM2020}. In particular, the latter solution was found to be essentially an extension to \replaced{that of}{the one by} \citet{Kessler1986}. With the addition of a small downflow, these two solutions are connected through an imperfect bifurcation, in which the transcritical bifurcation point subsequently evolves into a saddle-node point. 
\replaced{Furthermore}{Futhermore}, in \citet{JFM2020}, the existence of a steady solution was found to be limited within a certain range of the parameters. Indeed, for a sufficiently large downflow, the solution obtained with the GTD model does not exist when the Richardson number, which would depict the averaged cell-number density (see (\ref{eq:non_dim_para})), is greater than a critical value.

In \deleted{contrast, in }this study, we will extend the finding of \citet{JFM2020} by seeking solutions at sufficiently low flow rate\added{. In contrast to \citet{JFM2020}, here we seek}	\deleted{, where }solutions \deleted{exist }at Richardson number higher than the threshold. In particular, we \added{will }report that countably \replaced{infinitely many}{infinite number of} solutions exist due to the coupling between the flow and cell-number density equations. We will show that all these solutions emerge through a sequence of transcritical bifurcations, as the Richardson number increases.

\section{Problem formulation\label{sec:GovEq}}
Following the same assumptions as \cite{Bearon2012}, \cite{Croze2013,Croze2017} and \cite{JFM2020}, in the present study, we are only concerned with the axisymmetric and axially-uniform solution of the suspension in an infinitely long vertical pipe. The flow variables, therefore, consist of the streamwise (downward) velocity $U(r;t)$ and cell concentration $N(r;t)$ as a function of the radial position $r$ and the time $t$. Following \cite{JFM2020}, the equations of motion are given by
\begin{subequations} \label{eq:base_full}
    \begin{equation}\label{eq:U0}
        \pardt{U}=- G+\frac{1}{Re}\frac{1}{r}\pardr{} \added{\left(r \pardr{U} \right)}+Ri (N-1),
    \end{equation}
    \begin{equation}\label{eq:n0}
        \pardt{N}=-\frac{1}{r}\pardr{}(r N \esth{r})+\frac{1}{D_R}\frac{1}{r}\pardr{}(r {D}_{rr} \pardr{N}),
    \end{equation}
with the boundary conditions and compatibility conditions:
    \begin{equation} \label{eq:vel0_n0_BC}
        U(1) = 0, \quad  \left[\esth{r} N-\frac{{D}_{rr}}{D_R}\pardr{N} \right] \Big|_{r=1} = 0, \quad
       \left. \pardr{U} \right|_{r=0} = 0, \left. \quad \pardr{N} \right|_{r=0} = 0.
    \end{equation}
\end{subequations}
Here, the equations have been non-dimensionalised by the radius of the pipe $h^*$, the cell swimming speed $V_s^*$ and the averaged cell concentration $N^*$. The dimensionless parameters of interest are
\begin{eqnarray}
\Ri=\ddfrac{N^* \upsilon^* g'^*h^*}{V_c^{*2}}, &
\Rey=\ddfrac{V_c^*h^*}{\nu^*}, &
\lambda=\ddfrac{1}{2 B^* D_R^*}, ~
D_R=\frac{D_R^* h^*}{V_c^*},\label{eq:non_dim_para}
\end{eqnarray}
where $\Rey$ is the Reynolds number, $\Ri$ the Richardson number, $\lambda$ the inverse of gyrotactic time scale normalised by rotational diffusivity  and $D_R$ the dimensionless rotational diffusivity, sometimes also known as the swimming Peclet number. Here, $\nu^*$ is the \replaced{kinematic }{fluid }viscosity \added{of the fluid}, $\upsilon^*$ the cell volume, $g'^*$ the reduced gravity, $B^*$ the gyrotactic time scale and $D_R^*$ the \deleted{(modelled) }isotropic rotational diffusivity, \added{which models the randomness in the rotational motion of swimmers.}
Also, $G=\added{\partial p/\partial z -Ri}$ is the dimensionless driving pressure gradient that excludes the hydrostatic pressure from the negative buoyant cells. As mentioned in \S\ref{sec:introduction}, the net cell swimming $\esth{r}$ and effective cell diffusivity $D_{rr}$ in the radial direction are modelled using the GTD theory. Hence, $\esth{r}$ and $D_{rr}$ are functions of the local shear rate $S=-\partial_r U /D_R$, the strength of gyrotaxis ($\lambda$) and rotational diffusion ($D_R$). The shear rate $S$ is sometimes known as the rotary P\'{e}clet number \citep{Hinch1972a}. 
Since the total number of the cells is preserved over the given control volume, we impose the normalisation condition for the cell concentration,
\begin{equation}\label{eq:n_norm}
    \int_0^1 N(r;t) r dr=\frac{1}{2}.
\end{equation}
Unlike a stationary suspension, the current set up allows for a net axial flow rate. Therefore, there is an extra degree of freedom which can be specified either by the pressure gradient $G$ or by the flow rate $Q$ given by
\begin{equation}\label{eq:Q}
    \int_0^1 U(r;t) r dr=\frac{Q}{2 \pi}.
\end{equation}
\added{If $Q$ is prescribed, then the pressure gradient is obtained as $G=(2/Re) \partial_r U|_{r=1}$.}

Lastly, the steady version of (\ref{eq:base_full}) can be further simplified into
\begin{subequations}\label{eq:base_steady}
\begin{equation}\label{eq:Nr_explicit}
    N(r)= N(0) \exp \Big(D_R \int_0^r \frac{\esth{r}}{D_{rr}} dr\Big),
\end{equation}
and
\begin{equation}\label{eq:U_G}
    -\frac{1}{Re}\mathcal{D}^2{U}= Ri \left(N(0) \exp \left(D_R \int_0^r \frac{\esth{r}}{D_{rr}} dr\right)-1\right)- G,
\end{equation}
\end{subequations}
where $N(0)$ is determined by (\ref{eq:n_norm}) and $G$ is determined by (\ref{eq:Q}). Here, we denote the radial Laplace operator by $\mathcal{D}^2=(1/r) \partial_r (r \partial_r)$.

\section{Numerical solutions \label{sec:Numerics}}
Following the same approach as \citet[\S 2.6]{JFM2020}, we solve for the steady solution to (\ref{eq:base_full}) numerically\deleted{ using Newton-Ralphson method}, where $r$\added{-dependence} is discretised using a Chebyshev collocation method. \added{After that, the resulting algebaric equations are solved using a Newton-Raphson method}. The computation is performed with \added{the number of nodes }$N_r=250$\replaced{. Numerical convergence is achieved as the numerical results only differ from}{, showing no difference from} \replaced{those of}{the results with} $N_r=175$ \added{by $0.2\%$ on average}.
We then perform \added{a }pseudo-arclength \replaced{continuation}{continuations} of the solutions by varying $Ri$ but at \added{a} fixed flow rate $Q$. The values of all parameters are taken from Table 2 of \cite{JFM2020}, which shall not be repeated here for brevity.
\begin{figure}
	\centering{}
	\sidesubfloat[]{\includegraphics[width = 0.96 \columnwidth]{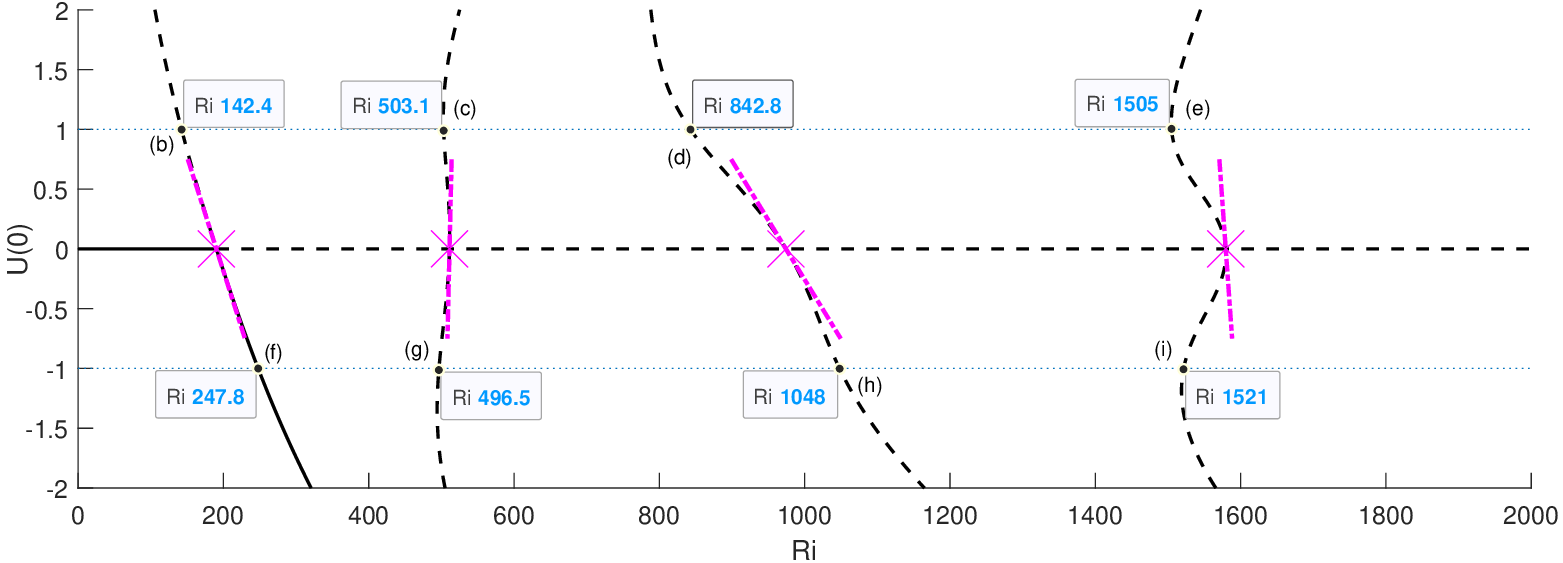}\label{fig:GTD_U0_bi}}\\
	\sidesubfloat[]{\includegraphics[scale = 0.4]{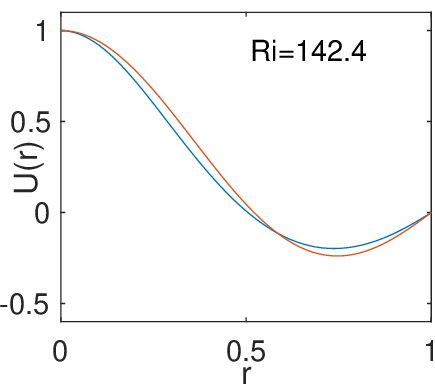}}\,
	\sidesubfloat[]{\includegraphics[scale = 0.4]{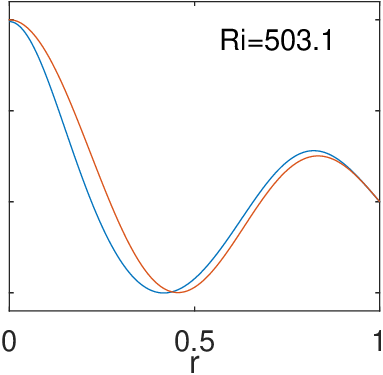}}\,
	\sidesubfloat[]{\includegraphics[scale = 0.4]{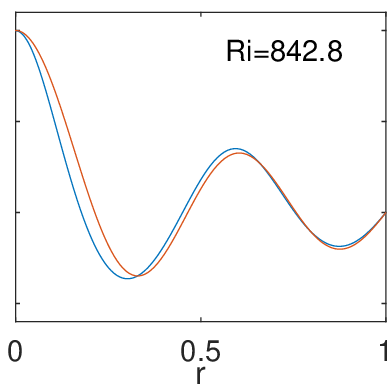}}\,
	\sidesubfloat[]{\includegraphics[scale = 0.4]{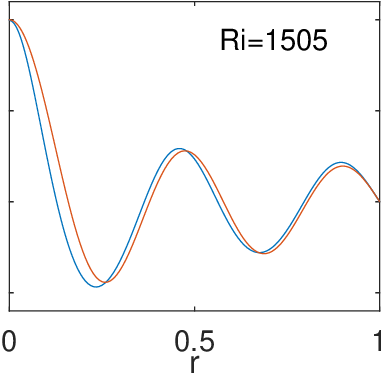}}\\
	\sidesubfloat[]{\includegraphics[scale = 0.4]{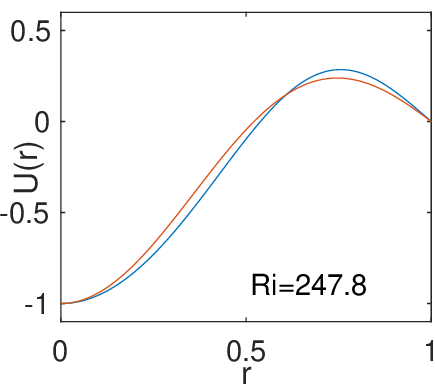}}\,
	\sidesubfloat[]{\includegraphics[scale = 0.4]{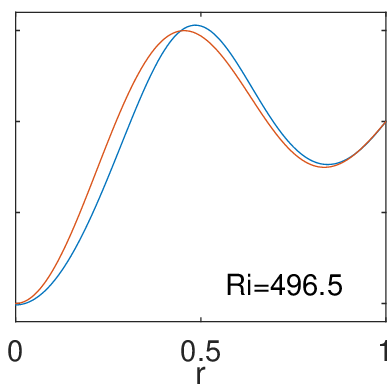}}\,
	\sidesubfloat[]{\includegraphics[scale = 0.4]{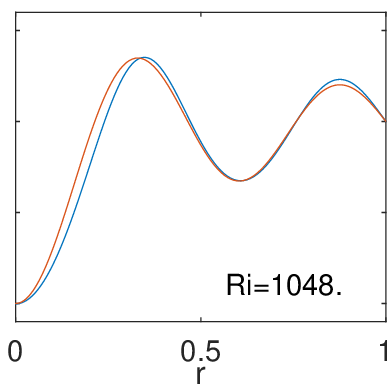}}\,
 	\sidesubfloat[]{\includegraphics[scale = 0.4]{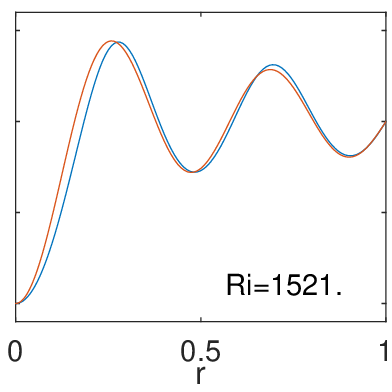}}\\
	\caption{\added{Comparison between numerical solutions of (\ref{eq:base_steady}) and the linear and weakly non-linear analysis in \S \ref{sec:4}. 
 $(a)$ Bifurcation diagram ($U(0)$ against $Ri$) of the $Q=0$ solutions from  (\ref{eq:base_steady}). Here, the line type represents the stability of the solution: \protect \solid, stable; \protect \dashed, unstable. The bifurcation point ($Ri_{c,n}$) from (\ref{eq:Ri_c}) is denoted by magenta crosses (x), while the slope of the bifurcation curve calculated from (\ref{eq:amp_eq}) is indicated by the short magenta dot-dash segment (\protect \dashdot).
 The dotted blue lines (\protect \dotted), representing $U(0)=\pm 1$, are added to locate the $Ri$ values for ($b$-$i$).  
 ($b$-$i$) Comparisons between the nonlinear steady solution (blue lines) and the corresponding (appropriately normalised) linear instability mode $\hat{u}(r)$ at $Q=0$ and $Ri$ as indicated in $(a)$. 
}}\label{fig:GTD_U0}
\end{figure}
\begin{figure}
	\centering{}
	\includegraphics[width = 0.96 \columnwidth]{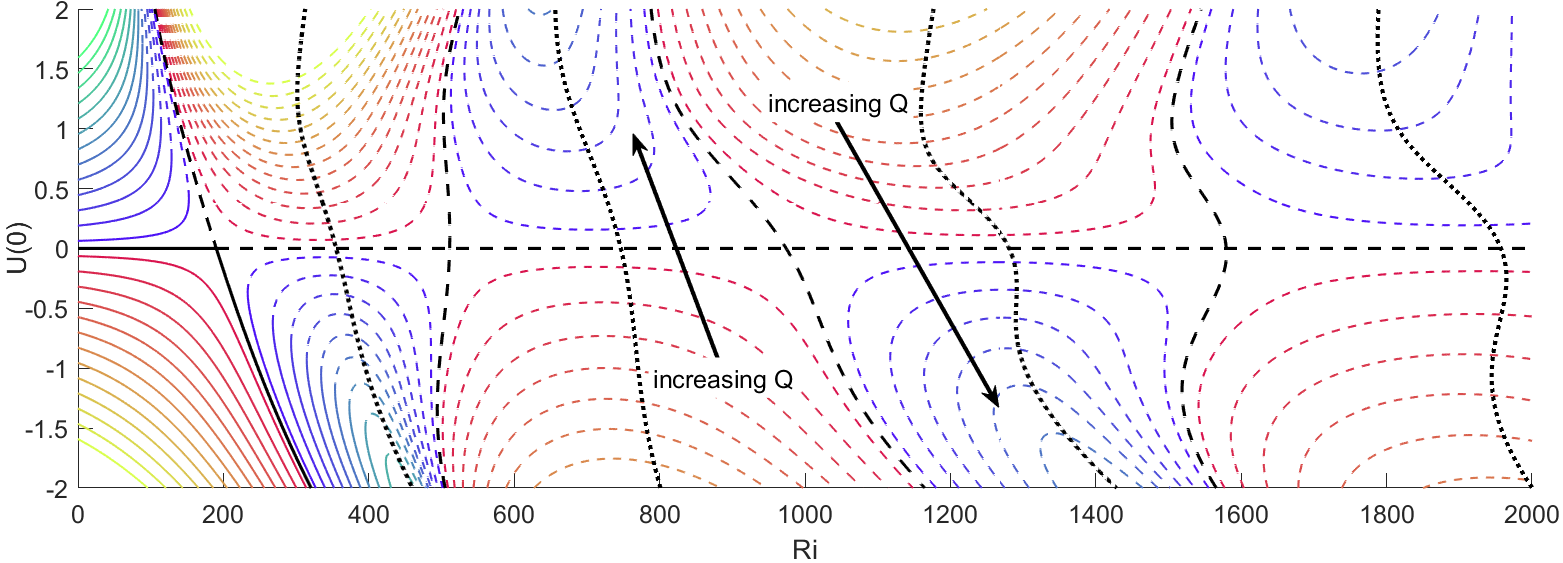}\\ 
	\caption{\added{
	Contour of $U(0)$ against $Ri$ at each given $Q \in [-2.5,2.5]$.
	Here, each contourline in the $U(0)-Ri$ plane indicates the value of $Q$: black, $Q=0$; blue to green, $Q \in [0.1,2.5]$ with $0.2$ increment; red to yellow, $Q \in [-2.5,-0.1]$ with $-0.2$ increment, while the line type represents the stability of the solution: \protect \solid, stable; \protect \dashed, unstable. Meanwhile, the thick dotted black lines (\protect \dotted)  shows the bifurcation by varying $Q$ for given $G=0$ (see \S \ref{sec:Qc}). 
	}}\label{fig:GTD_U0_bi_Q}
\end{figure}
In \citet[\S 3.1 and figure 2]{JFM2020}, we have briefly shown \added{that }there exist multiple branches of solutions at each $Q$ as long as $Q$ is near zero. Here, the same bifurcation diagram, but with a larger range of $Ri$ and $Q$, is presented\deleted{ in figure 1}. \added{In figures} \ref{fig:GTD_U0_bi} \added{and \ref{fig:GTD_U0_bi_Q}, }\deleted{where }the axial velocity $U(0)$ is used to represent the state of the steady solutions. \added{Figure \ref{fig:GTD_U0_bi} shows how the solution from (\ref{eq:base_steady}) changes with $Ri$ at $Q=0$, while figure \ref{fig:GTD_U0_bi_Q} shows the same at other $Q$ in the range $[-2.5,2.5]$.}
We have also plotted some of the steady solutions $U(r)$ when $|U(0)|=1$ and $Q=0$ in figure \ref{fig:GTD_U0}($b$-$i$, blue lines), for reasons that would become apparent later in \S\ref{subsec:linear}. The respective value of $Ri$ for each $U(r)$ can be found in each subfigure of figure \ref{fig:GTD_U0}\added{($b$-$i$)}, \added{as well as figure}  \ref{fig:GTD_U0_bi}.

Focusing on $Q=0$, from figure \ref{fig:GTD_U0_bi}, it is apparent that there exist multiple branches of steady solutions other than the uniform-suspension solution represented by $U(0)=0$. In fact, \added{as we increase $Ri$, there seems to be }a countable but infinite number of branches\deleted{appear to}  \replaced{emerging}{emerge} via \added{a sequence of }transcritical bifurcation\added{s}\deleted{, as $\Ri$ is increased}. Each branch coincides with the uniform suspension \replaced{at}{sate} a certain $Ri$, which forms the bifurcation point. \added{We shall define $Ri_{c,n}$ as the Richardson number of each bifurcation point from the left in figure \ref{fig:GTD_U0_bi}, where $n$ is the index. For convenience, we shall also index the $Q=0$ branches accordingly too. }\deleted{we shall number the solution branches using the $Q=0$ branches and their bifurcations, and call the one experiencing. }\added{For example, }the \added{first }bifurcation at \deleted{the lowest }$Ri(=Ri_{c,1} \approx 190)$ \added{is connected to }the first \added{$Q=0$ }branch (see the black line crossing the left most $Ri_c$ in figure \ref{fig:GTD_U0_bi}). At each bifurcation point, there is also an exchange of stability, as we examine the stability of the solution near each bifurcation (see \S\ref{subsec:linear}). However, except for the first bifurcation, the stability exchange takes place not with the most unstable mode but with a less unstable one. 
When a small downflow or upflow is applied (i.e. $Q \neq 0$), each transcritical bifurcation point turns into a saddle-node point via an imperfect bifurcation. This is similar to the finding of \cite{JFM2020} for the first branch solution, but the same happens to all the other branches.

\section{The origin of \added{an} infinite number of solution branches \label{sec:4}}
In this section, we will examine the linear stability of a uniform suspension in the vertical pipe. We will further restrict the perturbation to be axisymmetric, parallel and axially uniform, given the nature of the solutions of interest. \added{To be consistent with our numerical results in \S \ref{sec:Numerics} and figure \ref{fig:GTD_U0_bi}, in this section we will fix the flow rate when adding the perturbation, i.e. $Q=0$.} We will demonstrate that the multiple transcritical bifurcations of the computed solutions are the result of the pipe geometry, which also confines the suspension in a \deleted{finite }domain \added{with finite horizontal extent}. We will then extend it to \added{the }weakly nonlinear regime, similarly to previous work by \citet{Bees1999}. The resulting amplitude equation demonstrates that all the bifurcations in figure \ref{fig:GTD_U0_bi} are indeed transcritical.

\subsection{Linear stability analysis\label{subsec:linear}}
We first consider a perturbation to the stationary uniform suspension in a cylindrical pipe with infinite depth: i.e.  $U=\epsilon u_1(r,t)+\epsilon^2 u_2(r,t)+O(\epsilon^3)$ and $N=1+\epsilon n_1(r,t)+\epsilon^2 n_2(r,t)+O(\epsilon^3)$. Given that $\esth{r}(S)$ is odd and $D_{rr}(S)$ is even with respect to the radial shear rate $S$, this yields. 
\begin{equation} \label{eq:GTD_lin_model}
    \esth{r}=\epsilon\beta \pardr{u_1}+\epsilon^2\beta \pardr{u_2}+\mathcal{O}(\epsilon^3), \quad \mbox{and} \quad \frac{D_{rr}}{D_R}=D+\left.\frac{\epsilon^2}{2 D_R^3}\frac{\partial^2 D_{rr}}{\partial S^2}\right|_{S=0} \left(\pardr{u_1}\right)^2+\mathcal{O}(\epsilon^3),
\end{equation}
where $\beta=-(\partial_S \esth{r}|_{S=0})/D_R$ and $D=D_{rr}|_{S=0}/D_R$ are constants that depend only on $D_R$ and $\lambda$.

We also note that $\beta/D=\lambda/2$ \citep[see][]{Bearon2012}. 
At $\mathcal{O}(\epsilon)$, the perturbed equations for linear stability are then obtained as:
\begin{subequations} \label{eq:lin_full}
    \begin{equation}\label{eq:up}
        \pardt{u_1}=\frac{1}{Re}\mathcal{D}^2 u_1+Ri~n_1- {G_1},
    \end{equation}
    \begin{equation}\label{eq:np}
        \pardt{n_1}=-\beta \mathcal{D}^2 u_1+D \mathcal{D}^2 n_1,
    \end{equation}
with boundary conditions
    \begin{equation} \label{eq:lin_BC}
        u_1(1) = 0, \quad  \left[\beta \pardr{u_1}-D\pardr{n_1} \right] \Big|_{r=1} = 0.
    \end{equation}
\replaced{Because we have fixed $Q=0$,}{and} the perturbed pressure gradient $G_1$ is found such that \added{the flow rate is not altered by the perturbation, i.e. }
    \begin{equation}\label{eq:up_int}
      \int_0^1 u_1 r dr=0.
    \end{equation}
\end{subequations}

While (\ref{eq:lin_full}) can be solved numerically, we shall proceed to focus on the special cases when one of the stability modes is neutrally stable. Introducing a neutrally stable and stationary normal mode (i.e. $u_1(r,t)=\hat{u}(r)$ and $n_1(r,t)=\hat{n}(r)$), (\ref{eq:lin_full}) is then simplified into a single equation:
\begin{equation}\label{eq:4.3}
    \mathcal{D}^2 \hat{u} +\kappa^2 \hat{u}=G_1,
\end{equation}
where $\kappa^2= RiRe\beta/D= RiRe\lambda/2$. The left-hand side of (\ref{eq:4.3}) is the Bessel differential equation, which admits the Fourier–Bessel series as solutions. \replaced{Substituting}{With} the boundary conditions, the mode shapes \added{of} \replaced{$\hat{u}(r)$}{$u_1(r)$} \added{should take} \deleted{takes} the form\deleted{ of}
\begin{equation}\label{eq:Bessel_ur}
    \hat{u}(r)=\frac{G_1}{\kappa^2}\left( 1- \frac{J_0(\kappa r)}{J_0(\kappa)} \right),
\end{equation}
where $J_m(r)$ is the $m^{th}$ Bessel function of \added{the }first kind. Enforcement of (\ref{eq:up_int}) into (\ref{eq:Bessel_ur}) subsequently leads to an infinite number of discrete values of $\kappa(=\kappa_{c\added{,n}})$ satisfying $J_2(\kappa_{c,n})=0$, at which (\ref{eq:Bessel_ur}) becomes a neutrally stable solution to (\ref{eq:lin_full}). \added{Here, $n$ indicates the $n^{th}$ zeros of $J_2(r)$.} In this case, $G_1$ in (\ref{eq:Bessel_ur}) becomes an arbitrary real constant, as (\ref{eq:up_int}) is satisfied for any $G_1$. The values of $\kappa_{c\added{,n}}$ also yield the critical values of 
\begin{equation}\label{eq:Ri_c}
  Ri_{c\added{,n}}=2\kappa_{c\added{,n}}^2/(Re \lambda)
\end{equation}
for neutral stability of each mode. These values of $Ri_{c\added{,n}}$ \added{calculated from $\kappa_{c,n}$} match perfectly with the bifurcation points \added{computed numerically in the previous section}, as shown in figure \ref{fig:GTD_U0_bi}. Finally, it should be mentioned that (\ref{eq:Ri_c}) is equivalent to (3.14) in \cite{Pedley1988} and (31) in \cite{Bees1999} where a continuous set of the critical values of the parameters equivalent to $Ri$ and $\kappa$ are obtained from linear stability analysis. However, in the present study, the introduction of a finite domain in the radial direction results in discrete values of $\kappa_{c\added{,n}}$ and $Ri_{c\added{,n}}$ with the corresponding eigenmode in the form of a cylindrical harmonic (i.e. Bessel functions) that satisfies the given boundary conditions.

\subsection{Weakly nonlinear analysis \label{subsec:weakly}}

We further proceed to perform a weakly nonlinear analysis close to $Ri_{c\added{,n}}$. In the previous study by \cite{Bees1999} where \added{a} uniform suspension is considered in \added{an} unbounded domain, it was assumed that the leading nonlinear term would appear at the third order due to translational invariance of the suspension in the horizontal direction. However, in the present study, such invariance is broken due to the pipe's cylindrical geometry. Hence,  there is no reason that the leading nonlinear term would emerge at the third order. In this study, we therefore start by assuming $Ri-Ri_{c\added{,n}}=\epsilon \Delta Ri$ ($\Delta Ri$ is the normalised distance from the bifurcation point) with a slow time scale $T=\epsilon t$. Given the perturbation form introduced in \S\ref{subsec:linear}, at $\mathcal{O}(\epsilon^2)$, we get
\begin{subequations} \label{eq:O2_pert}
\begin{equation}
    \partial_T u_1-\Delta Ri~n_1  = -\partial_t u_2-G_2 + \frac{1}{Re} \mathcal{D}^2 u_2 +Ri_{c\added{,n}}~n_2,
\end{equation}{}
where $G_2=(2/Re) u_2'(1)$ and
\begin{equation}
\partial_T n_1+\beta \frac{1}{r}\partial_r \left( r(\partial_r u_1) n_1 \right)  = -\partial_t n_2-\beta \mathcal{D}^2 u_2 + D \mathcal{D}^2n_2.
\end{equation}
\end{subequations}
We also introduce amplitude $A(T)$ for the linear perturbation:
\begin{equation}
    u_1(r,t,T)=A(T)\hat{u}(r),\quad n_1(r,t,T)=A(T)\hat{n}(r),
\end{equation}
where the linear instability mode is normalised to be $\hat{u}(0)=1$.
Following the procedure of weakly nonlinear analysis \cite[e.g.][]{Drazin-WeaklyNonlinear}, we apply the solvability condition to (\ref{eq:O2_pert}) using the adjoint of (\ref{eq:lin_full}). After some simplifications, we arrive at the amplitude equation
\begin{equation}
    \left( C-\frac{F}{D Re} \right)\partial_T A- \frac{\lambda}{2} C \Delta  Ri A -\frac{\lambda E}{2 Re} A^2=0, \label{eq:amp_eq}
\end{equation}
where $C$, $E$ and $F$ are defined in appendix \ref{app:nonlinear}.
Now, given the quadratic nonlinearity in (\ref{eq:amp_eq}), it is clear that all the bifurcations in figure \ref{fig:GTD_U0_bi} are transcritical. We have also computed the slopes of the non-trivial branches in figure \ref{fig:GTD_U0_bi}, which are given by $-C Re/E$, when they cross each bifurcation point. As shown by the dot-dashed lines in figure \ref{fig:GTD_U0_bi}, the computed slopes match with those of the numerically computed nonlinear solutions perfectly at the bifurcation points.

Finally, the neutrally stable eigenmodes (\ref{eq:Bessel_ur}) used for the weakly nonlinear analysis (red line, with appropriate sign) are compared with the numerical solutions (blue line) \replaced{around}{at} each bifurcation point \added{(at $U(0)=\pm 1$)} in figure \ref{fig:GTD_U0}. As expected, there is excellent agreement between them. 

\section{Existence of steady solution \label{sec:Qc}}
\begin{figure}
		\centering{}
	\sidesubfloat[]{\includegraphics[width = 0.45 \columnwidth]{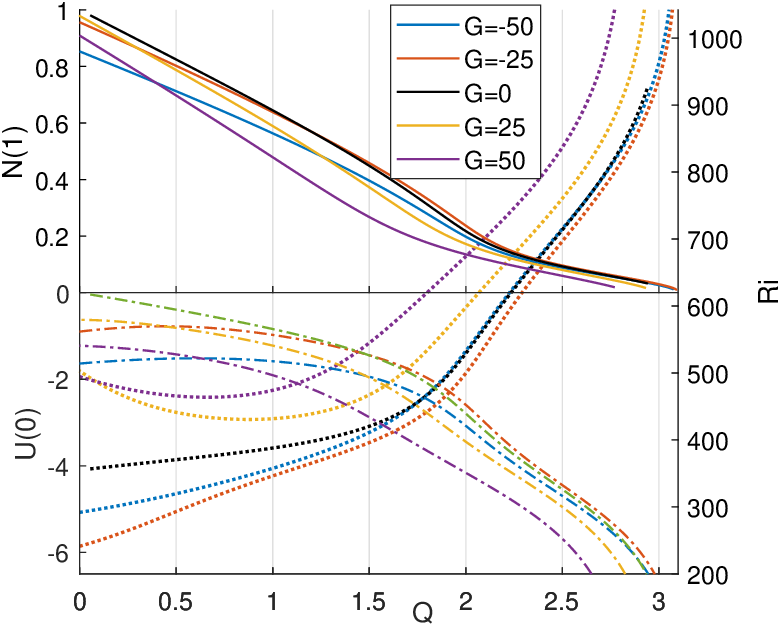}\label{fig:GTD_G00}}
	\sidesubfloat[]{\includegraphics[width = 0.45 \columnwidth]{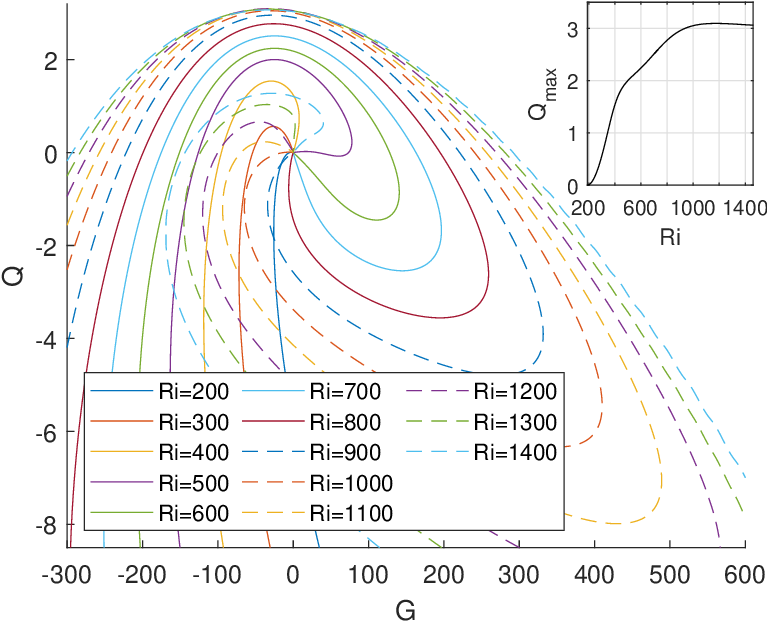}\label{fig:GTD_fixedRi}}
	\caption{Continuations of the steady solution emerged from the first bifurcation point with $U(0)<0$. $(a)$ $U(0)$(\protect \dashdot) , $Ri$(\protect \dotted) and $N(1)$(\protect \solid) for several $G$ on increasing $Q$ from $0$. $(b)$ The relation between $G$ and $Q$ for several fixed $Ri$. Here, in the inset, the maximum achievable flow rate $Q_{\max}$ is plotted for each $Ri$.  \label{fig:contG}}
\end{figure}

As previously noted by \cite{HP2014b} and \cite{JFM2020}, there are three parameters $Q$, $G$ and $Ri$ which control the bifurcation of (\ref{eq:base_full}). Here, $Q$ and $G$ are dependent on each other, providing only two degrees of freedom in total. Now, without loss of generality, we shall vary the three parameters $Q$, $G$ and $Ri$ in a controlled manner to explore the existence of the steady solutions reported in \S\ref{sec:Numerics}. We first prescribe $G$ and continue the steady solutions to (\ref{eq:base_full}) by changing $Q$ -- the related bifurcation diagrams for $G=0$ are also shown in figure \ref{fig:GTD_U0_bi_Q} (dotted lines). Here, we note that the change of $Q$ for a given $G$ requires a change of $Ri$, given the relation of the three parameters. Figure \ref{fig:GTD_G00} shows how $U(0)$, $Ri$ and $N(1)$ changes with $Q$ for several prescribed $G$. It is found that all the solutions blow up at a certain respective threshold of $Q$ (say $Q_{c,G}(G)$): as $Q \rightarrow Q_{c,G}(G)$, $U(0)$ and $Ri$ blow up and \replaced{$N(1)$}{$N(0)$} approaches zero (i.e. depletion of the cell number density at the wall).

To further explain the existence of a threshold value of $Q$, we also perform continuation by changing $G$ for prescribed $Ri$. Figure \ref{fig:GTD_fixedRi} shows the continuation of the first two steady-solution branches reported in figure \ref{fig:GTD_U0_bi} for fixed $Ri\in[200,1400]$. Now, it becomes apparent that there exists a maximum achievable $Q(\equiv Q_{\max,Ri}(Ri))$ at each $Ri$. Plotting $Q_{\max,Ri}(Ri)$ against $Ri$ (inset in figure \ref{fig:GTD_fixedRi}) shows that $Q_{\max,Ri}$ reaches its maximum at $Q_{\max} \approx 3.1$ and $Ri \approx 1174$. The maximum downward flow rate $Q_{\max}(>0)$ is typically achieved with a downward pressure gradient ($G<0$) (i.e. $Q_{\max}=\max{(Q_{c,G}(G))}$ in figure \ref{fig:GTD_G00}). However, any further decrease of $G$ \deleted{from that value }counter-intuitively \replaced{decreases}{makes} the flow rate of the steady solution \replaced{rather than increases}{decreased instead of being increased}. \deleted{Furthermore, while an increase of the averaged cell concentration (i.e. $Ri$) can also increase\deleted{s} \added{the} flow rate, there is still a maximum flow rate at certain $Ri$ value.}

Lastly, we note that figure \ref{fig:contG} is only for the first and second solution branches emerged from $Q=0$ in figure \ref{fig:GTD_U0_bi}. However, the same qualitative behaviours have been found from all the other solution branches: for example, the relation between $N(1)$ and $Q$, the blow up of $U(0)$ and $Ri$ as $Q\rightarrow Q_{c,G}$, and the existence of $Q_{\max}$. 

\section{Conclusion and discussion}
In this study, we have sought \added{for the} nonlinear, steady and axisymmetric solutions for a suspension \added{of gyrotactic swimmers} in an infinitely long pipe. \replaced{An}{A countably} infinite number of steady solutions have been found\replaced{. Each}{, and each} of them stems from a transcritical bifurcation on the uniform \replaced{solution}{suspension}. The \added{exact values of the} bifurcation points have also been found by solving the linearised equations for neutral stability\deleted{ analytically}.

Comparing the present study with \cite{Bees1999}, who showed the existence of \added{an} \deleted{`}uncountably\deleted{'} infinite number of solutions to a similar set of equations in a unbounded domain, we can conclude that the \deleted{`}countably\deleted{'} infinite number of transcritical bifurcations originate from the finite horizontal domain and the flow geometry (i.e. pipe).
Firstly, the finite horizontal domain yields discrete eigenvalues \replaced{from}{of} the equations for \added{the} linear stability (\ref{eq:lin_full}), making $(\kappa_{c\added{,n}},Ri_{c\added{,n}})$ a discrete set rather than a continuous curve. Hence we have a countably infinite number of bifurcation points.
Secondly, the cylindrical geometry of the pipe breaks \added{the }translational invariance in the horizontal direction. In \added{an} unbounded domain, such invariance of uniform suspension is broken by the primary bifurcation \cite[i.e. pitchfork bifurcation; see][]{Bees1999}. However, in the pipe, the primary bifurcation takes place in a circumstance where the translational invariance is already broken by the flow geometry, \replaced{which}{and this} leads the primary bifurcation to be transcritical \added{instead}. 

The existence of many steady solutions has also hinted at the possible dynamical route from a uniform suspension to the gyrotactic pattern. Except for the first branch, all the other steady-solution branches found in figure \ref{fig:GTD_U0_bi} are saddles in the state space. In other words, if a stationary and uniform suspension at $Ri>Ri_{c\added{,n}}$ is perturbed with the \added{$n^{th}$ most} linear\added{ly} \replaced{unstable}{instability} mode \replaced{from}{in} (\ref{eq:Bessel_ur}) \deleted{with $\kappa=\kappa_c$}, the system would first evolve towards the corresponding nonlinear steady solution. Therefore, the flow patterns related to the unstable solutions in the present study may well be observed at least transiently, before further development of the flow state or its breakdown in the axial and/or azimuthal directions. Indeed, early numerical simulations by \cite{Ghorai1999,Ghorai2000b} found such a transient dynamics, which strongly hinted at the dynamical importance of the initial perturbation. \added{Furthermore, the increasing number of nonlinear steady solutions and linearly unstable mode as $Ri$ increases strongly hinted at the increasing complexity of the system as $Ri$ increases.}

Finally, the emergence of multiple axisymmetric steady solutions with increasing $Ri$ implies that \replaced{similar}{such} solutions may well exist for non-axisymmetric case. This implies that the route to the final flow pattern would be a highly complicated process, involving competitions between the axisymmetric and non-axisymmetric states. Furthermore, it should also \replaced{be}{been} pointed out that all these axially uniform steady solutions are unstable to axially varying perturbations \cite[]{JFM2020}. To this end, there is an ongoing investigation to address these issue. 

\section*{Acknowledgement}
This work is funded by the President's PhD Scholarship of Imperial College London.

\section*{Declaration of interests}
The authors report no conflict of interest.

\appendix
\section{Weakly nonlinear analysis \label{app:nonlinear}}
Here, the full expressions for $C$, $E$ and $F$ in \S \ref{subsec:weakly} are derived. We define
\begin{equation}
f(r)=\left (1-\frac{J_0(\kappa_c r)}{J_0(\kappa_c)} \right) / \left( 1-\frac{1}{J_0(\kappa_c)}\right),
\end{equation}
such that the normalised first-order velocity profile is  $\hat{u}=f(r)$. The adjoint of $\hat{u}$ is the same as $\hat{u}$. Now, $\hat{n}=\lambda \hat{u}/2$, but the adjoint of $\hat{n}$ is $-(\beta Re)^{-1}g(r)$, where $g(r)$ is
\begin{equation}
g(r)= \left (J_0(\kappa_c r)-J_0(\kappa_c)+\frac{r^2-1}{2}\kappa_c J_1(\kappa_c) \right) / \left( J_0(\kappa_c)-1\right).
\end{equation}
From the solvability condition, we arrive at (\ref{eq:amp_eq}), in which the constant $C$, $E$ and $F$ are dependent on $\kappa_c$ and are defined by 
\begin{equation}
    C=\int_0^1f(r)^2rdr=\frac{\kappa_c^2}{8}\left( 1-\frac{1}{J_0(\kappa_c)}\right)^{-2},
\end{equation}{}
\begin{equation}
    E = \int_0^1 g(r) \partial_r\left[rf'(r)f(r)\right] dr,
\end{equation}{}
\begin{equation}
    F=\int_0^1f(r)g(r)rdr=-\frac{-3 \kappa_c^2}{16} \left( 1-\frac{1}{J_0(\kappa_c)}\right)^{-2}.
\end{equation}{}

While $C$ and $F$ are analytically found, $E$ is numerically integrated owing to the complexity of the Bessel function. The numerical values of $-C Re/E$ are plotted as the local slopes of the corresponding branch at each bifurcation point in figure \ref{fig:GTD_U0_bi}.

\bibliographystyle{jfm}
\bibliography{references}

\end{document}